\documentclass[twocolumn,prb,showpacs,preprintnumbers,superscriptaddress,amsmath,amssymb]{revtex4}
\usepackage{graphicx}
\usepackage{dcolumn}
\usepackage{color}
\usepackage{bm}
\newcommand{\dpsi}{\psi^\dagger}
\newcommand{\npsi}{\psi^{\phantom{\dagger}}}

\newcommand{\ud}{\,\mathrm{d}}

\begin{document}

\title{Transport and scattering in inhomogeneous quantum wires}

\author{N.~Sedlmayr}
\email{sedlmayr@physik.uni-kl.de}
\affiliation{Department of Physics, University of Kaiserslautern,
D-67663 Kaiserslautern, Germany}
\author{J.~Ohst}
\affiliation{Department of Physics, University of Kaiserslautern,
D-67663 Kaiserslautern, Germany}
\author{I. Affleck}
\affiliation{Department of Physics and Astronomy, The University of British Columbia,
Vancouver, BC V6T 1Z1, Canada}
 \author{J.~Sirker}
\affiliation{Department of Physics, University of Kaiserslautern,
D-67663 Kaiserslautern, Germany}
\affiliation{Research Center OPTIMAS, University of Kaiserslautern,
D-67663 Kaiserslautern, Germany}
\author{S.~Eggert}
\affiliation{Department of Physics, University of Kaiserslautern,
D-67663 Kaiserslautern, Germany}
\affiliation{Research Center OPTIMAS, University of Kaiserslautern,
D-67663 Kaiserslautern, Germany}
\date{\today}

\begin{abstract}
  We consider scattering and transport in interacting quantum wires
  that are connected to leads.  Such a setup can be represented by a
  minimal model of interacting fermions with
  sudden changes in interaction strength and/or velocity.  The
  inhomogeneities generally cause relevant backscattering, so it is {\it a
  priori} unclear if {perfect ballistic transport is
    possible in the low temperature limit. We demonstrate that a
    conducting fixed point surprisingly exists even} for large abrupt
  changes, which in the considered model corresponds to a velocity
  matching condition.  The general position dependent Green's function
  is calculated in the presence of a sudden change, and is confirmed
  numerically with high accuracy.  The exact form of the interference
  pattern in the form of density oscillations around inhomogeneities
  can be used to estimate the effective strength of local
  backscattering sources, {offering a route to design
    experiments where the effects of the contacts are minimized}.
\end{abstract}


\maketitle
The description of transport through quantum wires has become a very
well studied research area, as it directly ties together conductivity
experiments\cite{carbon,yacoby,LeHur1,tarucha} with central theoretical
models in one dimension, such as the Landauer formalism\cite{landauer} and
Luttinger liquids.\cite{Giamarchi}  Landauer showed that the conductance
of a single spinless non-interacting quantum channel is always
finite and given by $G= (1-R^2) e^2/h$, where $R$ is the
backscattering amplitude.\cite{landauer}  However, interaction
effects change this picture dramatically as $R$ becomes temperature
dependent through renormalization.  In the pioneering work of Kane and
Fisher it was found that a local perturbation in a repulsive Luttinger
liquid is relevant, which results in a
characteristic powerlaw dependence $R\propto T^{g-1}$ where $g<1$ is the
interaction parameter.\cite{PhysRevB.46.15233}  A number of impurity models in
one dimension have since been analyzed in detail
\cite{frojdh,qin,erik,2CK,sedlmayr,PhysRevB.62.4370,PhysRevB.46.10866,yue1994,Nagaosa}
and confirm that generically a local perturbation cuts the transport
at low temperatures, unless the relevant operator is forbidden by
symmetry.

Nearly perfect connections between leads and wires
can be achieved for different types of quantum wires,\cite{carbon,yacoby,LeHur1,tarucha}
which show quantized conductance at moderate temperatures.
At lower temperatures, however, scattering in the connections to the leads
plays an increasingly important role. The higher
dimensional contacts are effectively weakly interacting, so there
has been great interest in describing
the transport through a quantum wire attached to non-interacting
leads.\cite{ogata,pono,Maslov1995,rechprl2008,PhysRevB.52.R17040,PhysRevB.54.R5239,yue1994,chamon1997,wong1994,gutman2010,PhysRevB.66.035313,Enss,thomale}
The minimal model for this setup is a
single channel of interacting spinless fermions, where the interaction parameter
changes along the wire.  Under the assumption that
it is possible to use a hydrodynamic Luttinger liquid
description for this model,\cite{Maslov1995}
it would be expected that backscattering follows a non-trivial
renormalization behavior which is position dependent.\cite{PhysRevB.54.R5239}
Even if the connections are effectively free from imperfections
with a homogeneous lattice structure,\cite{tarucha} there have to be
small regions of the wire, the junctions, where the interaction changes, which does
not necessarily occur adiabatically and will induce intrinsic backscattering.
This immediately invites the question if it is ever possible
to create a perfect connection
in the low temperature limit.  Indeed it is in general unclear how large
such intrinsic backscattering is for generic junctions and if it is even
justified to use an inhomogeneous Luttinger liquid description in the first place
for this setup, since  the usual bosonization procedure assumes a translational and
scale invariant theory.

In this paper we consider the intrinsic backscattering in an inhomogeneous
wire and show that
 a perfectly conducting fixed point can indeed be found by adjusting
parameters such as the velocities on both sides. At half filling it is also
possible to estimate the bare strength of the backscattering from
the change in the local velocity field.
The full correlation function of an inhomogeneous Luttinger liquid is calculated,
which agrees with numerical simulations to high accuracy.  This confirms the validity
of the inhomogeneous theory and proves that a
low energy conducting fixed point can exist even in the presence of large abrupt jumps.

A hydrodynamic description of interacting fermions with a changing interaction parameter
$g_x$ leads to
a generalized inhomogeneous Luttinger liquid action\cite{Maslov1995,PhysRevB.54.R5239}
\begin{eqnarray}
S_0=\int_0^{1/T}\ud\tau\int\ud x \frac{u_x}{2 g_x}\left[
\frac{(\partial_\tau\phi)^2}{u^2_x}+(\partial_x\phi)^2\right], \label{action}
\end{eqnarray}
where $\phi$ is a canonical bosonic field.   Here the effective velocity $u_x$
generally also changes with the interaction strength.
We will derive the corresponding correlation function for an abrupt jump below.
Additionally, however,
backscattering must be considered.
In particular, it is known that the Hamiltonian density contains a
leading relevant oscillating operator given by\cite{PhysRevB.46.15233}
\begin{equation}
e^{- i 2 k_F x} \dpsi_+\npsi_- \propto e^{-i2  k_F x}
e^{-i \sqrt{4 \pi} \phi}, \label{relevant}
\end{equation}
where $\npsi_\pm$ are left and right moving fermion fields and $k_F$ is the Fermi-wavenumber.
Normally this operator can be neglected under the integral, but this is no longer the case
for inhomogeneous systems.
In fact, it is exactly this operator as a local perturbation which causes
the renormalization of defects in wires\cite{PhysRevB.46.15233} and
spin chains.\cite{PhysRevB.46.10866}
The oscillating part of the interaction
results in the same bosonic operator,\cite{PhysRevB.46.10866} so that
even a change in interaction alone will induce backscattering.

{In principle conductance depends on the connections of the
  wire to both of the leads. We will concentrate here on the backscattering in
the case where the length of the wire is larger than the coherence
  length $\propto u/T$ so that it suffices to consider each junction
  separately.\cite{PhysRevB.54.R5239}} A single junction can be
characterized by changes taking place in a small region around $x=0$.
The field $\phi_x$ is assumed to be slowly varying on the scale of the
Fermi wavelength so that it is possible to use an expansion of
$\phi_x$ in Eq.~(\ref{relevant}) and a partial integration to derive
an effectively local perturbation
\begin{equation}
  H' 
  \approx  \lambda e^{-i\sqrt{4\pi}\phi_{x=0}} + h.c.,
\label{oscillating}
\end{equation}
while the oscillating operator cancels everywhere in the uniform regions.
So far we have kept only the leading relevant operator, but there are higher order
terms which will be discussed later.
In general $\lambda$ is complex, except for particle hole symmetric situations.
Even relatively smooth
junctions become effectively sharper and sharper under renormalization, so that
the relevant backscattering is non-zero unless the amplitude
$\lambda$ is adjusted to zero, which requires the fine
tuning of two parameters.
As we will see later it is indeed possible
to identify such conducting fixed points in a lattice model by
an appropriate choice of parameters on both sides of the junction.
The existence of a conducting fixed
point is also of relevance for the discussion about possible charge
fractionalization in Luttinger liquids.\cite{Pham,LeHur1,LeHur2} If
the Luttinger liquid supports only charges $e(1\pm g_x)/2$
related to the chiral eigenstates of the Hamiltonian on each side of
the wire, then it seems to be impossible that backscattering can be
tuned to zero. Our results thus support the analysis in
Ref.~[\onlinecite{LeHur2}] that at such a junction an {\it arbitrary}
charge can be injected into a Luttinger liquid.

In order to perform a renormalization group analysis we have to consider the
full partition function
\begin{eqnarray}\label{partition}
\mathcal{Z}=\int D\phi e^{-S_0-\int_0^{1/T}\!\!\ud\tau H'}
\end{eqnarray}
with the quadratic action in Eq.~(\ref{action}).  For an abrupt change in the
interaction parameter at $x=0$
from $g_{x<0}= g_\ell$ to $g_{x\geq 0}=g_r$ and similarly for the velocity $u_x$
the general bosonic Green's function
$\tilde{G}(x,x';\tau)=\langle\phi_{x,\tau}\phi_{x',0}\rangle_0$ is
found by matching the left and right parts.  In particular, it is
possible to solve
\begin{equation}
\tilde{G}(x,x';\tau)=T\sum_me^{i\omega_m\tau}\tilde{G}_m(x,x') 
\end{equation}
with
\begin{equation}
\bigg[\frac{ \omega_m^2}{2 g_x u_x}-\frac{\partial}{\partial x}
\bigg(\frac{u_x}{2 g_x}\frac{\partial}{\partial x}\bigg)\bigg]\tilde{G}_m(x,x')
 =\delta(x-x')
\end{equation}
by allowing a discontinuity in the derivative of $\tilde{G}_m$ at $x=x'$.\cite{Maslov1995}
We determine the general Green's function to be
\begin{eqnarray} \label{GF}
\tilde{G}(x,x';\tau)&=&
-\frac{\bar g}{\pi}\ln\left|\sinh\left[\pi T\left(\frac{|x|}{u_x}
+\frac{|x'|}{u_{x'}}-i\tau\right)\right]\right|  \\
& + & \frac{\mathcal{L}[x,x']g_x}{\pi}\ln\left|\frac{\sinh\left[\pi T\left(\frac{|x|}{u_x}+\frac{|x
'|}{u_{x'}}-i\tau\right)\right]}{
\sinh\left[\pi T\left(\frac{|x-x'|}{u_x}-i\tau\right)\right]}\right|, \nonumber
\end{eqnarray}
where we have defined a new effective interaction parameter
${\bar g}= 2(\frac{1}{g_\ell}+\frac{1}{g_r})^{-1}$.
Here $\mathcal{L}[x,x']$ is $1$ when $x$ and $x'$ are in the same region,
and $0$ when they are not.
The renormalization of a local perturbation
in Eq.~(\ref{oscillating}) can be determined with the help of the Green's function by
integrating out the Fourier components above
a cutoff $\Lambda$,\cite{Giamarchi,PhysRevB.54.R5239} which gives
\begin{eqnarray} \label{RG}
\frac{1}{\lambda}\frac{d\lambda}{d\ln{\Lambda}}&=&1-{\bar g}.
\end{eqnarray}
We therefore expect that the effective backscattering renormalizes as
a power law in the temperature $R\propto T^{\bar g-1}$.

As a concrete lattice model we can
consider spinless fermions at half-filling
\begin{equation}
\label{microH}
 H= \sum_{x}   \left[-t_x(\psi_x^\dagger \psi_{x+1}^{\phantom{\dagger}} +h.c.)
+ U_x
(n_x-\tfrac{1}{2})(n_{x+1}-\tfrac{1}{2})\right],
\end{equation}
where $n_x=\psi^\dagger_x \psi_x^{\phantom{\dagger}}$.
The corresponding interaction parameter $g$ and the renormalized velocity $u$
in Eq.~(\ref{action})  are
functions of $U$ and $t$, which are known from the Bethe ansatz.\cite{Giamarchi}
For small jumps and interactions
we have estimated the size of $\lambda$ perturbatively,
which turns out to be proportional to
the corresponding renormalized velocity field $u_x$
\begin{eqnarray} \label{lambda}
\lambda  &\propto &\sum_x e^{-i2 k_F x }(u_{x+1}-u_x),
\end{eqnarray}
with $2 k_F=\pi$,
so to lowest order it does not matter if the velocity change occurs due to
inhomogeneous interactions or hopping amplitudes, which may even compensate each other.
In particular, Eq.~(\ref{lambda}) suggests that a conducting fixed point can be
achieved for a sharp discontinuity by a velocity matching
$u_\ell=u_r$, but the
derivation above is valid only for small interactions.
The numerical simulations show that this condition holds even in
a strongly interacting model.
Equation (\ref{lambda}) also implies that backscattering can be made arbitrarily small by
very slow "adiabatic'' changes.

In order to calculate the backscattering amplitude $R$ numerically,
we use quantum Monte Carlo simulations\cite{SSE} on systems
with an abrupt junction at $x=0$, i.e.~$U_{x<0} = U_\ell$ and $U_{x\geq 0} =
U_r$ and analogously for the hopping amplitude $t_x$.
For long system sizes $L \agt 40t_x/T$
the boundary condition at $\pm L/2$ becomes irrelevant.
The backscattering $R$ induces a $2k_F$ interference pattern in the density, the
so-called Friedel oscillations,\cite{yue1994,PhysRevB.62.4370,PhysRevLett.75.934}
 which can be
calculated directly in the simulations and also give additional information about the
correlation functions.
In particular, we consider the local oscillating density
in a half-filled
lattice 
in response to changing the chemical potential
\begin{eqnarray}
\chi_x&=&\left.\frac{\partial}{\partial \mu}\langle n_x\rangle\right|_{\mu=0},
\end{eqnarray}
where the density is bosonized as
\begin{eqnarray}
\label{density}
 n_x
& =&  n_0
-\frac{1}{\sqrt{\pi}}\partial_x\phi_x+\textrm{const.}
\frac{(-1)^{x}}{\pi}\sin[\sqrt{4\pi}\phi_x]
\end{eqnarray}
analogously  to the local magnetization and susceptibility in spin chains.\cite{PhysRevB.46.10866,PhysRevLett.75.934,PhysRevB.62.4370,SirkerLaflorencie,SirkerLaflorencie2,SirkerLaflorencieEPL}

\begin{figure}
\includegraphics*[width=0.8\columnwidth]{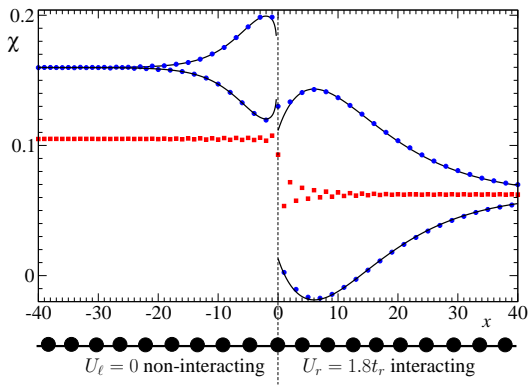}
\caption{(Color online) The local response in the density $\chi$ for a
  jump in interaction from $U_\ell=0$ to $U_r=1.8t_r$ at $T=0.1t_r$.
  Circles (blue): No discontinuity in hopping, $t_\ell=t_r$. Squares
  (red): The hopping on the left is adjusted to $t_\ell\approx
  1.518t_r$ in order to match the velocities on both sides.  Solid
  lines (black)
 are fits to the predicted behavior in
  Eq.~(\ref{lineshape}) for even and odd sites separately.}\label{oscil}
\end{figure}
The results for the Friedel oscillations $\chi=\chi_0+(-1)^x\chi_{\rm
  alt}$ near an abrupt junction at $x=0$ are shown in Fig.~\ref{oscil} for a change
from $U_\ell=0$ to $U_r=1.8t_r$ at finite temperatures $T=0.1t_r$ for two
cases: In the case that the interaction strength changes but the hopping is equal
$t_\ell=t_r$, we have $u_\ell < u_r$
 and strong alternating $2k_F=\pi$ oscillations are observed (circles).
In the second case, the hopping $t_\ell$ on the left was
also increased in order to exactly match the velocities $2 t_\ell=u_\ell=u_r
\approx 3.036t_r$ (squares).
Clearly, the
backscattering oscillation is strongly suppressed with a different
position dependence, but it is not zero.  As the numerical data will
show it turns out that the {\it relevant} backscattering term is
exactly zero in this case.  The uniform parts of $\chi_x$ on the two sides are
constant and approximately given by the zero temperature result
$\chi_0=\frac{g_x}{\pi u_x}$, which are not equal in either case.
We have also tested other cases with more complicated changes in
hopping and interactions over three sites in order to tune $\lambda=0$ in Eq.~(\ref{lambda}),
and in all cases backscattering is strongly
suppressed.

\begin{figure}
\includegraphics*[width=0.8\columnwidth]{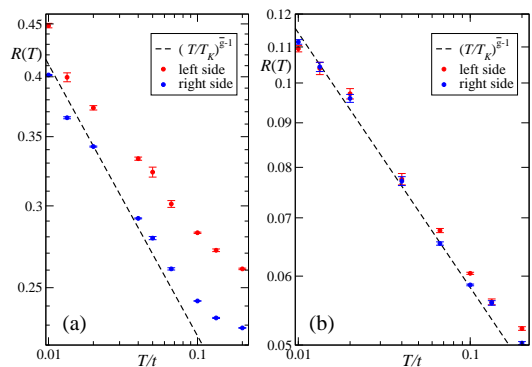}
\caption{(Color online) Effective backscattering amplitude $R(T)$ on a
  logarithmic scale extracted from the amplitude of the density
  oscillations in Eq.~(\ref{lineshape}) for a jump from (a) $U_\ell=0$
  to $U_r=1.8t$, ($T_K=3.4\times10^{-4}t$), and
(b) $U_\ell=1t$ to $U_r=1.4t$ with $t=t_\ell=t_r$, ($T_K=6\times10^{-6}t$).
  The amplitudes are extracted from fitting the local response for
  $x<0$ (``left side'') and $x>0$ (``right side'')
  separately.}\label{R}
\end{figure}

Let us first analyze the position dependence of the Friedel amplitude
in the case of unequal velocities,
which shows a characteristic maximum in Fig.~\ref{oscil}
reminiscent of the local susceptibility near ends in spin chains.\cite{PhysRevLett.75.934,SirkerLaflorencieEPL}  However, we will show
that the behavior is {\it not} exactly of the same form as for
scattering from open ends as was conjectured before.\cite{PhysRevB.62.4370}
  In order to calculate the alternating
response in the presence of the perturbation (\ref{oscillating}) with
small $|\lambda|\propto|u_r-u_\ell|$ we use the bosonized form of the
density (\ref{density})
\begin{eqnarray}
\chi_{\rm alt}\!&=& \!\lambda
\int_{0}^{1/T}\!\!\!\!\!\!\ud\tau
\frac{\partial}{\partial \mu}
\left\langle\sin\sqrt{4\pi}\phi_{x,\tau}
\cos\sqrt{4\pi}\phi_{0,0}
\right\rangle_0,
\end{eqnarray}
where the dependence on $\mu$ is given by a shift of the field $\partial_x \phi$
by $\frac{\mu g_x}{\sqrt{\pi}u_x}$ which turns the sine-dependence into a cosine-dependence with an additional factor of $x$.
Using the Green's function in Eq.~(\ref{GF}) we arrive at integrals of the form
$\int_0^1 d\tau |\sinh (X- i\pi \tau)|^{-2 g} = 2^g (\sinh 2X)^{-g} P_{-g}(\coth 2 X)$
where $P_{l}(z)$ is the Legendre function. Therefore we find
\begin{eqnarray} \label{lineshape}
\chi_{\rm alt}&\propto &
\lambda \frac{g_x x}{u_x}
\int_{0}^{1/T}\!\!\! \ud\tau
\left\langle \cos \sqrt{4 \pi}\phi_{x,\tau} \cos
\sqrt{4 \pi}\phi_{0,0}\right\rangle_0 \nonumber \\
& \propto & \lambda  T^{\bar g -1} x
\left(\frac{u_x}{T} \sinh\left[\frac{2\pi Tx}{u_x}\right]\right)^{-{g_x}}P_{-{\bar g}}(z)
\end{eqnarray}
where $z=\coth[2\pi Tx/u_x]$.
Compared to the scattering from an open boundary\cite{PhysRevLett.75.934,PhysRevB.62.4370}
there are two notable
differences: first of all there is an additional factor in the form of the
Legendre function, which changes the shape significantly near the
scatterer, but quickly approaches unity for $x\agt u_x/\pi T$.
Second, there is an additional temperature dependent factor $\propto
T^{\bar g-1}$ which is in agreement with the renormalization behavior
predicted in Eq.~(\ref{RG}).  As can be seen by the fit in
Fig.~\ref{oscil}, the behavior (\ref{lineshape}) describes the
numerical data perfectly, where $x$ is taken to be the position from
an effective scattering center.  We have confirmed the position
dependence for many different temperatures and discontinuities.  This
unambiguously shows that the action in Eq.~(\ref{action}) and the Green's function
(\ref{GF}) leading up to Eq.~(\ref{lineshape}) are a reliable description of the problem.

An
interesting detail in Fig.~\ref{oscil} is the maximum in the
non-interacting region where $U_\ell=0$ and $g_\ell=1$ which is absent
for an open end but now arises due to the Legendre function with $\bar
g <1$.  Physically this can be understood as a proximity effect, where
the behavior in a range on the non-interacting side is influenced by the
collective excitations on the other side.

The temperature dependent amplitudes of the fits to the
position-dependent part in Eq.~(\ref{lineshape}) directly give the
backscattering $R(T)$ in Fig.~\ref{R}, where we have used the
amplitude from open ends corresponding to $R=1$ as the normalization,
which is known exactly.\cite{SirkerLaflorencieEPL,Terras} The data
confirm the predicted power-law behavior at low temperatures.  Note
that the independent fits on both sides give roughly the same $R(T)$.
For the larger jump in the interaction $U=1.8t_r$ there are deviations
at higher temperatures coming from higher order operators (left
panel).  Below a characteristic temperature $T_K$ we expect the power
law to break down as the stable fixed point $R(0)=1$ is approached,
but this energy scale could not be reached in the simulations.

\begin{figure}
\includegraphics*[width=0.8\columnwidth]{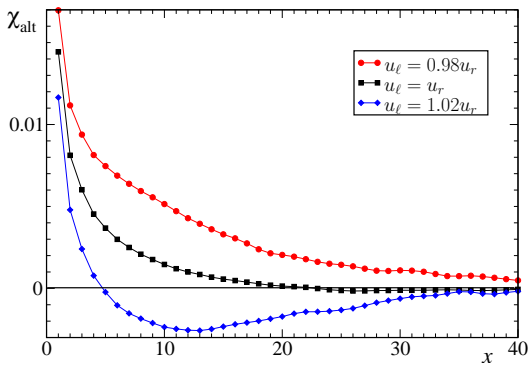}
\caption{(Color online) The alternating part of the density oscillations at $T=0.1t_r$ on the interacting side $U=1.8t_r$, where the hopping on the non-interacting side is
adjusted so that $u_\ell=0.98 u_r$,  $u_\ell= u_r$, and  $u_\ell=1.02 u_r$, respectively.
 }\label{equalv}
\end{figure}

In order to find a conducting scenario in the low temperature limit,
it is interesting to analyze the case of equal velocities $u_\ell=u_r$
with strong discontinuities in both hopping and interaction in more
detail.  As shown in Fig.~\ref{oscil} density oscillations are still
observed in this case, but we are interested only in the contribution
of the leading {\it relevant} operator in Eq.~(\ref{oscillating}),
which will grow while the temperature is lowered and must change sign as
a function of velocity at the conducting fixed point.  In
Fig.~\ref{equalv} we show the alternating part of the oscillations for
a jump from $U_\ell=0$ to $U_r=1.8t_r$, where the hopping on the left
has been adjusted to three different cases: $u_\ell=0.98 u_r$,
$u_\ell= u_r$, and $u_\ell=1.02 u_r$.  It is quite apparent that a
sign change takes place exactly at $u_\ell= u_r$, which means that the
relevant backscattering vanishes. The remaining oscillations visible
in Fig.~\ref{oscil} for $u_\ell=u_r$ are caused by higher order local
operators in Eq.~(\ref{oscillating}).  In particular, the next-leading
terms are given by $\partial_x e^{i\sqrt{4 \pi} \phi}$ and
$e^{i\sqrt{16 \pi} \phi}$ with scaling dimension $\bar{g}+1$ and $4
\bar{g}$, respectively, which are irrelevant unless $4
\bar{g}<1$. 
Away from half filling the marginal operator $\partial_x \phi_{x=0}$
will also be present,\cite{qin} but does not affect the scattering to
first order. The velocity matching rule for a conducting fixed point
is surprisingly simple, considering that it is not linked to any
special symmetry in this scenario. {Perfect conduction in
  quantum wires with impurities has been described before in cases
  where a renormalization to the periodic boundary condition fixed
  point occurs.\cite{PhysRevB.46.15233} This, however, is not the case
  here.  We find a novel conducting fixed point described by Eq.~\eqref{GF} which does not correspond to a standard boundary conformally invariant theory.}

To conclude, we have studied the intrinsic backscattering present when
connecting a quantum wire to a lead.  All inhomogeneities and
impurities in the junction are in general relevant for repulsive
Luttinger liquids, leading to a conductance which scales as
$G(T)\propto\frac{e^2}{h}[1-\left(T/T_K\right)^{2\bar{g}-2}]$, with an
unusual power-law exponent given by ${\bar g}=
2(\frac{1}{g_\ell}+\frac{1}{g_r})^{-1}$ in terms of the interaction
parameters on both sides of the junction $g_\ell,\ g_r$. We have shown
that it is possible to achieve a perfect connection, 
i.e.~an absence of relevant backscattering,
in inhomogeneous wires by tuning other
parameters such as the velocity in the lead and in the wire. The
general Green's function was calculated in the presence of an abrupt
jump along the wire based on an inhomogeneous Luttinger liquid action,
which was in excellent agreement with numerical simulations.  The
results suggest that the observation of Friedel oscillations in the
density along the wire, which can be attempted by scanning probe
microscopy, can { be used to analyze local scattering
  centers. A systematic study of inhomogeneities and thus an
  experimental test of our results seems feasible in semi-conductor
  heterostructures. One could start from a device where a quantum wire
  adiabatically broadens into a two-dimensional electron
  gas.\cite{tarucha} By applying a sharp potential barrier at the
  interface a local backscattering center could then be realized
  allowing it to continuously tune the backscattering amplitude
  $\lambda$.}

\section*{Acknowledgments}
We are thankful for discussions with J. Folk.
This work was supported by the DFG via the SFB/Transregio 49, NSERC, CIfAR, and the
MAINZ (MATCOR) graduate school of excellence.

\end{document}